# Laser-induced stress wave propagation through smooth and rough substrates

James D. Boyd[1], Martha E. Grady[1]*

[1]Department of Mechanical Engineering, University of Kentucky

*corresponding author email: m.grady@uky.edu

**Abstract**

We investigate laser-induced acoustic wave propagation through smooth and roughened titanium-coated glass substrates. Acoustic waves are generated in a controlled manner via the laser spallation technique. Surface displacements are measured during stress wave loading by alignment of a Michelson-type interferometer. A reflective coverslip panel facilitates capture of surface displacements during loading of as-received smooth and roughened specimens. Through interferometric experiments we extract the substrate stress profile at each laser fluence (energy per area). The shape and amplitude of the substrate stress profile is analyzed at each laser fluence. Peak substrate stress is averaged and compared between smooth specimens with reflective panel and rough specimens with reflective panel. The reflective panel is necessary because the surface roughness of the rough specimens precludes *in situ* interferometry. Through these experiments we determine that the surface roughness employed has no significant effect on substrate stress propagation and smooth substrates are an appropriate surrogate to determine stress wave loading amplitude of roughened surfaces less than 1.2 μm average roughness (Ra). No significant difference was observed when comparing the average peak amplitude and loading slope in the stress wave profile for the smooth and rough configurations at each fluence.

The laser spallation technique is used to characterize adhesion strength of a variety of thin film-substrate interfaces [1-5]. A key factor that often modulates adhesion strength within these systems is an increase in surface roughness [5-7]. While some researchers report an increase in adhesion strength [5, 6] as assessed by laser spallation, others report a decrease [7]. During these experiments, the effect of surface roughness on stress wave generation is often omitted. For example, Kandula *et al.* [5] studied the effects of increased surface roughness on adhesion of poly-p-phenylenebenzobisoxazole, and reported the largest root mean square surface roughness as 3.2 nm. While this magnitude of surface roughness might have little impact on acoustic wave propagation, the assumption that an increase in surface roughness by an order of magnitude, micron-sized roughness, should be addressed for applications where micron-sized roughness is expected.

Two such applications are orthopedic medical implants and thermal barrier coatings (TBCs). The successful integration and adhesion of osteoblastic cells to medical implants is vital for longevity and reduction of infections [8]. As such, the ability to accurately quantify the adhesion strength of osteoblastic cells and medical device surfaces is crucial to develop appropriate surfaces. The laser spallation technique has been employed to quantify the adhesion of biological systems, specifically cell adhesion [9]. TBCs are also a technology that benefits from increased adhesion with increased surface roughness [10]. While laser spallation techniques have been applied to adhesion measurement of electron-beam physical vapor deposited TBCs [11], surface roughness effects have not been examined.

The high energy Nd:YAG used to initiate spallation can be employed for a variety of systems, but to quantify the substrate stress profile generated from this laser, the free surface velocity must be obtained, typically using a Michelson-type interferometer [1]. Displacement measurements of this kind require a reflective surface, thus, roughened surfaces pose a problem as they refract the light needed to acquire a sufficient signal. Because of the adhesion effects of previously discussed surface roughness increases, it is important, for the expanded use of the laser spallation technique, to develop a system that allows for the continued use of a Michelson-type interferometer system to obtain substrate stress profiles. The significance of this work is twofold: (1) We have developed a method for laser spallation experimentalists to determine the effect of surface roughness on stress wave propagation and (2) we show that micron-sized surface roughness of titanium-coated glass has no effect on slope or amplitude of recorded stress wave profiles, thus widening the field of potential applications.

This work is motivated by an adhesion study of cells and biofilms on dental implant-mimicking titanium surfaces [6, 12]. Therefore, the chosen surfaces in this work represent smooth and rough commercial titanium dental implants. Scanning electron microscopy (SEM) images of smooth and rough substrates appear in **Fig 1**. The surface roughness achieved is an average of 1.2 µm Ra, a common value for dental implants [13, 14]. During preliminary calibration testing, we found that the roughened surface was not reflective, which precludes *in situ* interferometric signal capture during stress wave loading. To overcome the nonreflective nature of the roughened surfaces, we modified the substrate systems with a reflective panel.

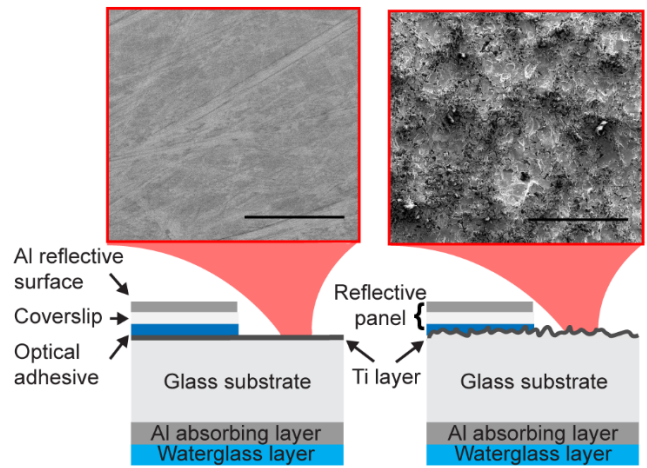

**Fig. 1.** Schematic of substrate assemblies with reflective panels. Components of the reflective panel are indicated on the left schematic. Thicknesses of component layers are not to scale. Representative SEM images of smooth (left) and rough (right) titanium surfaces are included. Scale bars on SEM images are 100 μm.

Substrate systems consist of 1 mm glass slides coated with a 100 nm layer of titanium, to mimic dental implant surfaces, and 300 nm aluminum as an energy absorption layer for the laser spallation process. Smooth and rough glass slides with titanium and aluminum layers were purchased from Deposition Research Laboratory, Inc. Surface roughness was confirmed by a Zygo white light interferometer. The aluminum layer is coated in optically clear sodium silicate to confine the compressive stress wave propagation towards the titanium-film interface, identical to substrates seen in Boyd *et al.* [15]. A reflective panel consists of a 170 μm thick coverslip (VWR micro cover glass No. 2), which is coated with 150 nm of aluminum by Lesker physical vapor deposition (PVD), and adhered to the surface with a thin layer, less than 5 μm, of Norland 60 optical adhesive. The reflective panel is applied to both the smooth and rough titanium samples. With both samples now reflective, the impact of the surface roughness on stress wave propagation is quantified and compared to the smooth sample substrate.

The substrate assemblies with reflective panels were inserted into our laser spallation system, illustrated in **Fig. 2**. A single pulse Nd: YAG laser is attenuated through a variable attenuator to control the energy of the laser. The laser pulse is then focused from an 8 mm diameter down to a 2 mm diameter to facilitate multiple loading sites on a single specimen. Each specimen is loaded approximately 5 times, and each loading condition is conducted on 2 substrate assemblies with reflective panels. The focused and energy-controlled pulse impinges upon the energy absorbing aluminum layer, initiating plasma gasification and converting the laser pulse into a compressive wave [16]. The compressive wave propagates through the substrate before arriving at the free surface. A Michelson-type interferometer measures the displacement of the free surface during loading and subsequent reflection of the wave. The interferometer includes a continuous wave solid state diode laser of 532 nm wavelength. The continuous laser, after collimation, passes through a beam splitter with half of the laser beam traveling to a fixed mirror and the other half traveling to the surface of the reflective panel. The continuous laser is aligned to the compressive stress loading location. The interferometer laser beams reflect off their respective surfaces and recombine where the interference pattern is incident upon a biased silicon photodetector (Electro Optics ET 2030). The photodetector is connected to a high-rate oscilloscope (LeCroy WaveRunner 8404 M), which captures the interference pattern at a sampling rate of 40 GS/s and transforms the change in light intensity into a voltage trace. Doppler equations are applied to the voltage trace to obtain surface displacement followed by constitutive equations to produce the substrate stress profile following previously established protocols [2, 5, 17, 18].

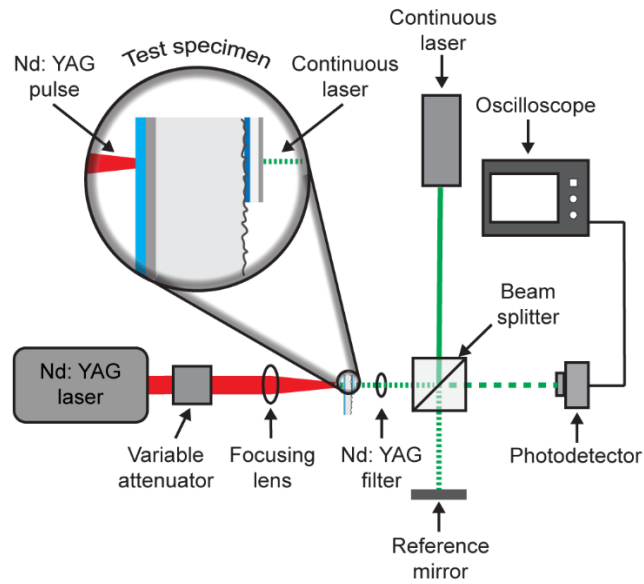

**Fig. 2.** Laser spallation set up used for experiments. The orientation of a substrate assembly with a reflective panel is shown with respect to Nd: YAG impingement and the continuous wave laser of the Michelson-type interferometer.

This sequence of steps is represented in **Fig. 3** by following the solid red line beginning in **Fig. 3(a)** where the voltage trace, V(t), is given by $V(t)=\frac{V_{max}+V_{min}}{2}+\frac{V_{max}-V_{min}}{2}*\sin(2\pi n(t))$, where $V_{max}$ and $V_{min}$ are the voltage maximum and minimum respectively of each interference fringe. The interference fringe number, n(t) is unwrapped and then converted to displacement (**Fig. 3(b)**) using $u(t) = \frac{\lambda_0 n(t)}{2}$, where $\lambda_0$ is the wavelength of the continuous wave laser, 532 nm. For a simple bi-material interface, the evolution of the substrate stress is readily determined from the displacement history using the principles of one-dimensional wave mechanics [2, 5]. The analytic thin film equation for the substrate stress, $\sigma_{sub}$ (**Fig. 3(c)**), is valid, $\sigma_{sub}(t) = -\frac{1}{2}(\rho C_d)_{sub}\frac{du}{dt}$ where $(\rho C_d)_{sub}$ denotes the density and dilatational wave speed of the substrate. For these experiments $\rho = 2500$ kg/m$^3$ and $C_d = 4540$ m/s are used as density and dilatational wave speed of glass, respectively. The largest magnitude of the compressive stress wave is called the peak substrate stress, which is averaged for each loading.

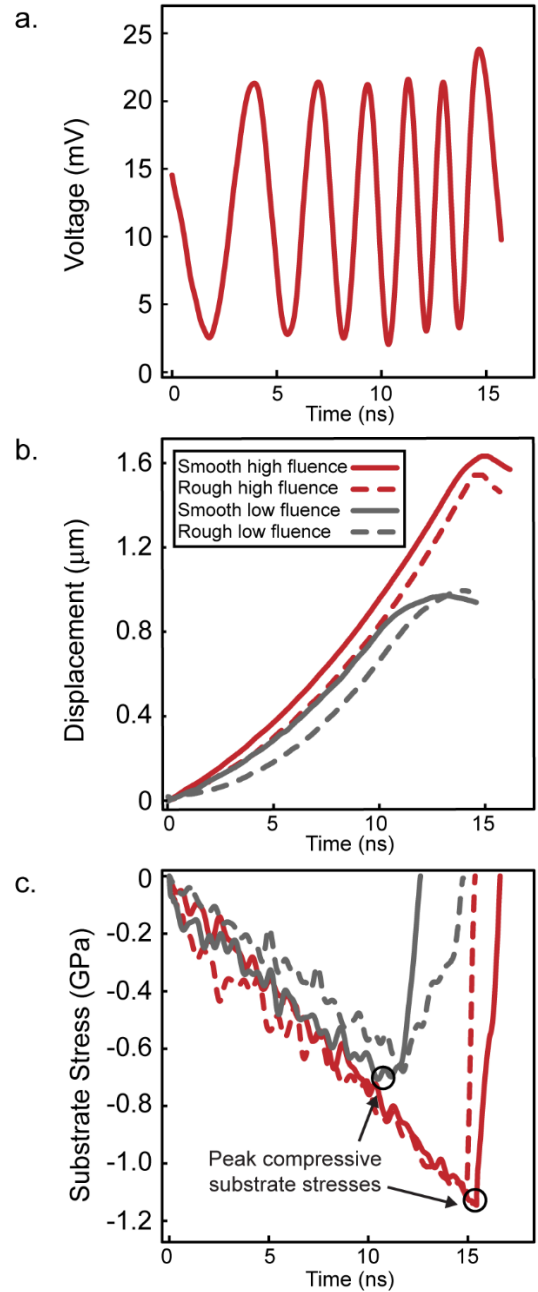

**Fig. 3. (a)** Representative voltage curve obtained from the Michelson-type interferometer under stress wave loading, **(b)** surface displacements, and **(c)** calculated substrate stress profiles. Peak compressive substrate stress is the maximum stress value recorded for each loading condition. Fluences of 79.4 mJ/mm$^2$ and 55.6 mJ/mm$^2$ are denoted by red and gray, respectively. Smooth and rough surfaces are denoted by solid and dashed lines, respectively. All interferometry data is collected on specimens with reflective panels.

Spallation experiments employ an increasing range of laser fluence values to accurately determine the minimum fluence needed to initiate ejection of the tested film. To determine the impact of surface roughness on the substrate stress profile, with the limited number of substrates with reflective panels created, effectively, two fixed fluences were examined. A high fluence, the maximum fluence with a 2 mm spot size, 79.4 mJ/mm$^2$, and a low fluence, 55.6 mJ/mm$^2$. These fluences were selected to ensure that fluence values critical to spallation were accounted for, and to ensure good repeatability with the limited samples.

Stress wave loading experiments were performed at the two fluences selected to calibrate free surface displacement to velocity as demonstrated in **Fig. 3**. The triangular shape of the compressive substrate stress profile is equable for both the smooth and rough calibration samples at the lower and higher fluence values and is consistent with the expected pulse shape in non-linear substrates like glass [1, 17]. The loading slopes across each fluence and substrate configuration aligned very well indicating the micron roughness has little impact on compressive stress wave propagation speed as expected.

The peak magnitude of the compressive substrate stress is likely to be more sensitive than slope or shape to surface modifications since these values rely primarily on the substrate material and not the interface. The peak compressive substrate stress value is averaged across 10 loaded regions for each substrate configuration at each fluence and is plotted in **Fig. 4**. The average ± standard deviation on smooth titanium at a loading fluence of 79.4 mJ/mm$^2$ is 1.16 ± 0.04 GPa, and on rough titanium is 1.14 ± 0.04 GPa. At the lower fluence of 55.6 mJ/mm$^2$, the average ± standard deviation on smooth titanium is 0.71 ± 0.07 GPa, and on rough titanium is 0.68 ± 0.05 GPa. It is important to examine peak compressive substrate stresses at each fluence with respect to the same surface type, as well as across different surface types, and the standard deviation of peak compressive substrate stress.

The average peak compressive substrate stress increases by 63% on the smooth substrates and by 68% on the rough substrates by increasing fluence from 55.6 mJ/mm$^2$ to 79.4 mJ/mm$^2$. As fluence increases, the imparted compressive stress wave magnitude is greater, and has been the mechanism by which laser spallation experimentalists have been able to determine adhesion strength of film-on-substrate systems [15, 17-20]. This important relationship is preserved with the reflective panel in place.

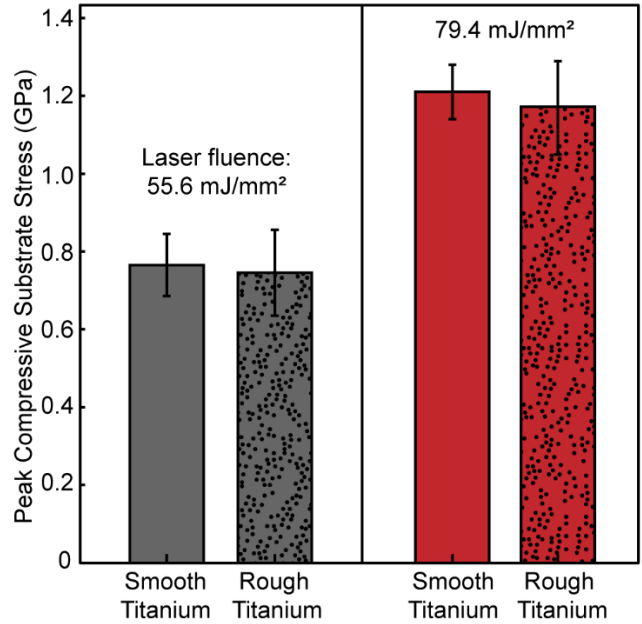

**Fig. 4.** Average peak compressive substrate stress obtained during loading at a fluence of 55.6 mJ/mm² (gray) and 79.4 mJ/mm² (red) on smooth (solid) surfaces and rough (dotted) surfaces. Error bars represent standard deviation.

The average of the peak compressive substrate stress at the low fluence (gray bars in **Fig. 4**) is nearly identical between the two different surfaces, which differ by 0.03 GPa, less than 3%. At the high fluence (red bars in **Fig. 4**), the result is the same; the average of the peak compressive substrate stress between rough and smooth surfaces differs by 0.02 GPa, less than 3%. A Student's t-test confirms the two data sets cannot be distinguished with p-values of 0.282 and 0.321 for fluences of 55.6 mJ/mm² and 79.4 mJ/mm², respectively. Because the peak compressive substrate stress values for the low and high fluences are indistinguishable based on surface roughness, the results obtained from calibrations on smooth substrates are an adequate substitution for calibrations on roughened substrates.

Analysis of the standard deviation of peak compressive substrate stress demonstrated strong repeatability across each fluence. Standard deviation ranged from 0.04 to 0.07 GPa, which corresponds to the variation in peak compressive stress reported by Grady *et al.* [17]. In that study, the standard deviation in peak compressive substrate stress is 0.02 GPa to 0.07 GPa, which was recorded on 1.5 mm thick fused silica substrates loaded at laser fluences between 14 and 50 mJ/mm². Further, the variation in peak compressive substrate stress does not seem dependent on substrate type as observed in another study on linear elastic silicon substrates, which had similar variance [3]. Thus, the standard deviation of peak compressive substrate stress in this study is not altered significantly by fluence or surface type, which supports rigor in sample preparation ensuring that the addition of adhesive and coverslips were all of consistent thickness.

The consistency in average peak compressive substrate stress across surface roughness for both fluences tested and the similarity in variation in peak compressive substrate stresses with other works, support the conclusion that the surface roughness used in this study has minimal impact on substrate stress wave propagation. This result means that the substrate stress profiles obtained on smooth samples can be used for calibration with the spallation data obtained on the roughened titanium, overcoming the challenge to collect interferometric data on a rough substrate [6]. Because of this finding, further research can easily be conducted into adhesion of biological films onto medical implant mimicking surfaces without the need for cumbersome calibration protocols. Similarly, TBCs, as stated previously, which often benefit from surface roughened metal surfaces to increase adhesion can be calibrated effectively, along with countless film/substrate systems which benefit from micron-sized surface roughness.

The results of this study are compelling to describe the effect of single micron surface roughness on laser-induced stress wave propagation. However, this study does not guarantee that surfaces with larger average roughness values would yield the same result. We anticipate at some critical surface roughness above single micron roughness that air will be trapped between film and substrate during film deposition due to high aspect ratio surface roughness. This trapped air between film and substrate will act to dissipate the stress wave. Thus, the comparison of adhesion measurements on high-aspect ratio rough substrates verse smooth substrates will not be an equal comparison as the amplitude of the stress wave will be attenuated. This work demonstrates that the threshold is above 1.2 μm Ra, which provides critical aid to experimentalists working in several key areas including medical devices and TBCs.

In summary, a straightforward protocol to determine the effect of surface modifications on laser-induced stress wave propagation has been developed. By comparing the peak compressive substrate stress values obtained for a smooth and rough surface we are able to determine that surface roughness values up to 1.2 μm Ra produce no measurable impact on stress wave propagation. While this relationship will likely not hold for increasing surface roughness values, the technique demonstrated in this work could be vital to future work that quantifies adhesion of films to rough substrates through the laser spallation technique.

**Acknowledgments**

We acknowledge NIH COBRE funding under grant number P20GM130456 for completion of these experiments. The project described was supported by the NIH National Center for Advancing Translational Sciences through grant number UL1TR001998. We also acknowledge Dr. Tom Berfield from University of Louisville as well as the UofL Micro/Nano Technology Center for guidance and use of their Lesker physical vapor deposition equipment.

**Data Availability**

The data that support the findings of this study are openly available in the Materials Data Facility [21, 22] at http://doi.org/10.18126/tw5w-xtwe, reference number [12].